\documentclass{article}
\usepackage{amsmath}
\usepackage{epsfig}
\usepackage{psfig}
\usepackage{latexsym}
\usepackage{amsfonts}
\usepackage{amssymb}
\usepackage{graphicx}
\begin{document}
\begin{flushright}
MZ-TH-05-12 \\
FTUV-05-07-19 \\
\end{flushright}

\medskip

\begin{center}
{\LARGE $D$ and $D_{S}$ decay constants from QCD duality at three
loops\footnote{{\LARGE {\footnotesize Supported by MCYT-FEDER under contract
FPA2002-00612 and partnership Mainz-Valencia Universities.}}}} \vspace{2cm}

{\large J. Bordes and J. Pe\~{n}arrocha}

{\normalsize Departamento de F\'{\i}sica Te\'{o}rica-IFIC, Universitat de
Valencia-CSIC}

{\normalsize E-46100 Burjassot-Valencia, Spain}

\vskip0.5cm

{\large K. Schilcher}

{\normalsize Institut f\"{u}r Physik, Johannes-Gutenberg-Universit\"{a}t}

{\normalsize D-55099 Mainz, Germany\vspace{2cm}}

\textbf{Abstract\medskip}
\end{center}

\begin{quote}
We compute the decay constants of the pseudoscalar mesons $\ D$ \ and
$\ D_{s}$ using a linear combination of finite energy sum rules which minimize
the contribution of the unknown continuum spectral function. We employ the
recent three loop calculation of the pseudoscalar two-point function expanded
in powers of the running charm quark mass. The theoretical uncertainties
arising from the QCD asymptotic expansion are quite relevant in this case due
to the relative small scale of the charm mass. We obtain the following
results: $f_{D}=177\pm21$ \ \ MeV and $f_{D_{s}}=205\pm22$ \ \ MeV. These
results, within the error bars, are in good agreement with estimates obtained
using Borel transform QCD sum rules, but somewhat smaller than results of
recent lattice computations.

\bigskip
\end{quote}

PACS: 12.38.Bx, 12.38.Lg. {\newpage}

\vspace*{2cm}

\section{Introduction}

The decay constant of a pseudoscalar meson $D_{q}$ consisting of a heavy
$c$-quark and a light quark $q=u,d,s$ is defined through the matrix element of
the corresponding pseudoscalar current as follows:
\[
<\Omega|\,(m_{c}+m_{q})\,\overline{q}\,i\,\gamma_{5}\,c(0)\,|D_{q}%
>\,=\,\,f_{D_{q}}\,M_{D_{q}}^{2}.
\]

Recently, two different experimental groups have extracted the decay constant
$\,f_{D^{+}}$ from \ a direct measurement of the absolute branching fraction
for the Cabbing-suppressed leptonic decay mode $D^{+}\rightarrow l^{+}\upsilon$
\cite{CLEO,BESII}. The results are quite different, although there is some
overlap of the large experimental errors. Both results are compatible with the
upper limit of $290$ MeV, established by the MARK-III collaboration at 90\%
C.L. \cite{MARKIII}. For the $D_{s}^{+}$ meson, there are several measurements
of \ its decay constant $f_{D^{+}}$published in the last decade
\cite{WA75,CLEOs,BESs,E653,BEATRIZ,OPAL}, with results in the range of
$194-430$ MeV. On the theoretical side, there is a recent estimate based on
Borel transformed sum rules $f_{D}=195\pm20$ MeV \cite{PENINSTEINHAUSER}, and
a preliminary result $\,f_{D}=225+11-13\pm21$ MeV \cite{LATTICE} obtained in
three flavor lattice QCD. There appears to be some room for improvement of the
accuracy of these results as well as for the systematic study of their
uncertainties. On the other hand, the result $f_{D_{s}}/f_{D}=1.18$ for the
ratio of the decay constants appears to be well established in lattice
QCD \cite{LATTICE,CP-PACS}.

In this letter, we estimate the decay constants $f_{D}$ and $f_{D_{s}}$ of the
peudoscalar charmed mesons using an alternative method, based on finite energy
sum rules, which compare moments of available experimental data with the
corresponding QCD theoretical counterpart. In particular, we take linear
combinations of positive moments in such a way that the contribution of the
data in the region above the resonances turns out to be practically
negligible. On the theoretical side we use a large momentum expansion of
massive QCD at three loops. This expansion is known up to the seventh power of
$m_{c}^{2}/s$, where $m_{c}$ is the mass of the charm quark and $s$ is the
square of the CM energy \cite{CHETYRKINSTEINHAUSER}. The expansion makes sense
as long as $s$ is far enough above the continuum threshold and above the
resonances. On the phenomenological side of the sum rule we consider only the
lowest lying pseudoscalar $D_{q}$-meson, once the unknown continuum
contribution has been removed by a judicious use of quark-hadron duality in
our method \cite{ACD}.

The plan of this note is the following. In section 2 we briefly
review the finite energy sum rule method employed. In section 3 we discuss the
theoretical and experimental inputs and we present our estimates for the decay
constants. Finally, in section 4 we write up the conclusions.

\section{The method}

The two point function associated with the pseudoscalar current is:
\begin{equation}
\Pi(s=q^{2})\,=\,i\int\,dx\,e^{iqx}<\Omega|\,T(j_{5}(x)\,j_{5}(0))\,|\Omega>,
\label{2PointFunc}%
\end{equation}
where $<\Omega\,|$ is the physical vacuum and the current $j_{5}(x)$ is the
divergence of the axial-vector current:
\begin{equation}
j_{5}(x)\,=\,(M_{Q}+m_{q}):\overline{q}(x)\,i\,\gamma_{5}\,Q(x):
\label{Current}%
\end{equation}
$M_{Q}$ is the mass of the heavy charm quark $Q(x)$ and $m_{q}$ is the mass of
the light quarks $q(x)$, up, down or strange. The starting point of our sum
rules is Cauchy's theorem applied to the two-point correlation function
$\Pi(s)$, weighted with a polynomial $P(s)$:
\begin{equation}
\frac{1}{2\pi i}\oint_{\Gamma}P(s)\,\,\Pi(s)\,\,ds=0. \label{Cauchy}%
\end{equation}

The integration contour $\Gamma$ extends over a circle of radius $\left\vert
s\right\vert =s_{0}$, and along both sides of the physical cut $\,\left[
s_{phys.},s_{0}\right]  $ where $s_{\text{\textrm{phys}}.}$ is the physical
threshold and $s_0$ is a duality parameter to be fixed with some stability criteria. 
The polynomial $P(s)$ does not change the analytical properties of
$\Pi(s)$. Therefore we obtain the following sum rule:
\begin{equation}
\frac{1}{\pi}\int_{s_{\text{\textrm{phys.}}}}^{s_{0}}\,P(s)\,\mathrm{Im}%
\,\Pi(s)\,\,ds=-\frac{1}{2\pi i}\oint_{\left\vert s\right\vert =s_{0}%
}\,\,P(s)\,\,\Pi(s)\,\,ds.\label{SR}%
\end{equation}

On the left hand side of equation (\ref{SR}) we use the experimental information
for $\mathrm{Im}\,\Pi^{\text{\textrm{DATA}}}(s)\,$between the physical
threshold $s_{\text{\textrm{phys}}.}$ and $s_{0}$, whereas on the right hand
side we use the asymptotic expansion $\Pi^{\mathrm{QCD}}(s)$ of QCD, including
perturbative and non-perturbative terms. The QCD expansion constitutes a good
approximation of the two-point correlate on the circle for a large enough
integration radius $s_{0}$.

The experimental data are dominated by the first pseudoscalar $c\overline{q}$
resonance. In the narrow width approximation, the absorptive part of the
two-point function $\mathrm{Im}\,\Pi^{\text{\textrm{DATA}}}(s)$ can be split
in two terms, the contribution of the resonance and the contribution of the
hadronic continuum $\mathrm{Im}\Pi^{\text{\textrm{cont}}}$ starting at the
physical continuum threshold $s_{\text{\textrm{cont}}.}>s_{\text{\textrm{phys.}}%
}=M_{D_{q}}^{2}$, as follows:%
\begin{equation}
\frac{1}{\pi}\,\,\mathrm{Im}\Pi^{\text{\textrm{DATA}}}(s)\,\,\,=\,\,\,M_{D_{q}%
}^{4}\,f_{D_{q}}^{2}\,\,\delta(s-M_{D_{q}}^{2})\,\,+\,\,\frac{1}{\pi
}\,\,\mathrm{Im} \, \Pi^{\text{\textrm{cont}}}\,\,\theta(s-s_{\text{\textrm{cont}%
}.}) \label{IMAG1}%
\end{equation}
where $M_{D\mathrm{_{q}}}$ and $f_{D\mathrm{_{q}}}$ are respectively the mass
and the decay constants of the lowest lying pseudoscalar meson $D_{q}$.

For the QCD correlate we write the decomposition,
\begin{equation}
\Pi^{\mathrm{QCD}}(s)\,=\,\Pi^{\mathrm{pert.}}(s)\,+\,\Pi^{\mathrm{nonpert.}%
}(s), \label{2QCD}%
\end{equation}
We employ the two-point correlation function $\Pi^{\text{\textrm{pert.}}}(s)$
with one massless and one heavy quark given to second order (three loops) in
the strong coupling constant $\alpha_{s}$ and expanded in a power series in
the pole mass of the heavy quark including terms of order $(M_{c}^{2}/s)^{7}$.
In \cite{CHETYRKINSTEINHAUSER} the following compact expansion of the
two-point function in terms of the pole mass $M_{c}$ can be found%
\begin{equation}
\Pi^{\text{\textrm{pert}}.}(s)=\Pi^{(0)}(s)+\left(  \frac{\alpha_{s}(M_{c}%
)}{\pi}\right)  \Pi^{(1)}(s)+\left(  \frac{\alpha_{s}(M_{c})}{\pi}\right)
^{2}\Pi^{(2)}(s), \label{PERT1}%
\end{equation}
where the different terms of the expansion in $\alpha_{s}$ have the form:
\begin{equation}
\Pi^{(i)}(s)=(M_{c}+m_{q})^{2}M_{c}^{2}\sum_{j=-1}^{6}\sum_{k=0}^{3}%
A_{j,k}^{(i)}\left(  \ln\frac{-s}{M_{c}^{2}}\right)  ^{k}\left(  \frac
{M_{c}^{2}}{s}\right)  ^{j}. \label{PERT2}%
\end{equation}
In the equations (\ref{PERT1},\ref{PERT2}), $M_{c}$ is the pole mass of the
charm quark. The coefficients $A_{j,k}^{(i)}$ are explicitly given in
\cite{CHETYRKINSTEINHAUSER}. For instance, the one-loop term of the expansion
in $\alpha_{s}$\ reads :
\begin{align*}
\Pi^{(0)}(s)\,  &  =\,\frac{3}{16\pi^{2}}(M_{c}+m_{q})^{2}s\left\{
3-2\log\left(  \frac{-s}{M_{c}^{2}}\right)  +4\frac{M_{c}^{2}}{s}\log\left(
\frac{-s}{M_{c}^{2}}\right)  \right. \\
&  -\left[  3+2\log\left(  \frac{-s}{M_{c}^{2}}\right)  \right]  \left(
\frac{M_{c}^{2}}{s}\right)  ^{2}+\frac{2}{3}\left(  \frac{M_{c}^{2}}%
{s}\right)  ^{3}+\frac{1}{6}\left(  \frac{M_{c}^{2}}{s}\right)  ^{4}\\
&  \left.  +\frac{1}{15}\left(  \frac{M_{c}^{2}}{s}\right)  ^{5}+\frac{1}%
{30}\left(  \frac{M_{c}^{2}}{s}\right)  ^{6}+\frac{2}{105}\left(  \frac
{M_{c}^{2}}{s}\right)  ^{7}+.....\right\}
\end{align*}

The non-perturbative terms in the asymptotic expansion of equation (\ref{2QCD})
are due to the quark and gluon condensates. We will include terms up to
dimension six \cite{REINDERS,NARISONQCD}:
\begin{align}
\Pi^{\mathrm{nonpert.}}(s)  &  =(M_{c}+m_{q})^{2}\left\{  M_{c}\left\langle
\bar{q}q\right\rangle \left[  \frac{1}{s-M_{c}^{2}}\left(  1+2\frac{\alpha
_{s}}{\pi}\right)  +2\frac{\alpha_{s}}{\pi}\ln\frac{M_{c}^{2}}{-s+M_{c}^{2}%
}\right]  \right. \nonumber\\
&  -\frac{1}{12}\frac{1}{s-M_{c}^{2}}\left\langle \frac{\alpha_{s}}{\pi}%
G^{2}\right\rangle -\frac{1}{2}M_{c}\left[  \frac{1}{(s-M_{c}^{2})^{2}}%
+\frac{M_{c}^{2}}{(s-M_{c}^{2})^{3}}\right]  \left\langle \overline{q}\sigma
Gq\right\rangle \nonumber\\
&  \left.  -\frac{8}{27}\pi\left[  \frac{2}{(s-M_{c}^{2})^{2}}+\frac{M_{c}%
^{2}}{(s-M_{c}^{2})^{3}}-\frac{M_{c}^{4}}{(s-M_{c}^{2})^{4}}\right]
\alpha_{s}\left\langle \overline{q}q\right\rangle ^{2}\right\}
\nonumber\label{NOPERT}%
\end{align}
For the quark condensate we include the $\alpha_{s}$ correction
\cite{JAMINLANGE}, it turns out to be small but non-negligible.

In order to improve the convergence of the perturbative expansion, we replace
the pole mass by the running mass using the $O(\alpha_{s}^{2})$ result
relating the two \cite{KS,CHETYRKINSTEINHAUSER1,MELNIKOV,VERMASEREN}. The
perturbative piece of order $(\alpha_{s})^{i}$ of equation (\ref{PERT2}) can be
rewritten in the form
\begin{equation}
\Pi^{(i)}(s)=m_{c}^{2}(\mu)(m_{c}(\mu)+m_{q}(\mu))^{2}\sum_{j=-1}^{6}%
\sum_{k=0}^{3}\tilde{A}_{j,k}^{(i)}\left(  \ln\frac{-s}{\mu^{2}}\right)
^{k}\left(  \frac{m_{c}^{2}(\mu)}{s}\right)  ^{j}\ \ \ \ \ (i=0,1,2).
\label{12}%
\end{equation}
and similarly for the non-perturbative piece. The coefficients $\tilde
{A}_{j,k}^{(i)}$ depend on the mass logarithms $\ln(m_{c}^{2}/\mu^{2})$ up to
the third power. As $\Pi(s)^{\rm QCD}$ is not known to all orders in $\alpha_{s}$, the
results of our analysis will depend to some extend on the choice of the
renormalization point $\mu$. In the sum rule considered here\ there are two
obvious choices, $\mu=m_{c}$ and $\mu=\sqrt{s_{0}}$. The former choice will sum up the mass logs of the form
$\ln(m_{c}^{2}/\mu^{2})$ and the latter choice the $\ln(-s/\mu^{2})$ terms.
For definiteness, we take $\mu=m_{c}$. With this, the convergence of the
perturbative terms is reasonably good. The results differ from taking
\ $\mu=\sqrt{s_{0}}$\ by an amount consistent with the general three-loop
asymptotic uncertainties, as we will analyze below.

Looking back to equation (\ref{SR}) and taking all the theoretical parameters
as well as the mass of the $D_{q}$-meson and the physical continuum threshold
as inputs of the calculation, we see that the decay constants can be computed
from equation (\ref{IMAG1}) only if we have good control over the hadron
continuum contribution of the experimental side.

To cope with this problem we make use of the freedom of choosing the
polynomial in equation (\ref{SR}). We take for $P(s)$ a polynomial of the
form:
\begin{equation}
P_{n}(s)=a_{0}+a_{1}s+a_{2}s^{2}+a_{3}s^{3}+\ldots+a_{n}s^{n}, \label{POLY}%
\end{equation}
where the coefficients are fixed by imposing a normalization condition at
threshold
\begin{equation}
P_{n}\left(  s_{\mathrm{phys.}}=M_{D_{q}}^{2}\right)  \,=\,1, \label{POLYNORM}%
\end{equation}
and requiring that the polynomial $P_{n}(s)$ should minimize the contribution
of the continuum in the range $\left[  s_{\mathrm{cont.}},s_{0}\right]  $ in a
least square sense, i.e.,
\begin{equation}
\int_{s_{\mathrm{cont.}}}^{s_{0}}s^{k}P_{n}(s)\,\,ds=0\,\,\mathrm{for}%
\,\,k=0,\ldots n-1, 
\label{POLY2}%
\end{equation}
The polynomials obtained in this way are closely related to the Legendre
polynomials. In the appendix the explicit form of the set of polynomials used
in this work is given.

This way of introducing the polynomial weight in the sum rule minimizes the
continuum contribution $\frac{1}{\pi}\,\,\mathrm{Im}\,\Pi^{\mathrm{cont.}%
}\,\,\theta(s-s_{\mathrm{cont.}})$ on the phenomenological side of the sum
rule. To the extend that $\mathrm{Im}\,\,\Pi^{\mathrm{cont.}}$ can be
approximated by an $n$-th degree polynomial these conditions lead actually to
an exact cancellation of the continuum contribution to the left hand side of
equation (\ref{SR}). The role of the $D_{q}$ resonance will be enhanced. We
will see in our analysis that this choice of the polynomial has the additional
effect of increasing the region of duality characterized by the value of the
duality parameter $s_{0}$. In this way the asymptotic expansion of QCD can be
used more safely on the circular integration contour. Notice however that
increasing the degree of the polynomial $P_{n}(s)$ will require the knowledge
of further terms in the mass expansion and in the non-perturbative series.
Therefore the polynomial degree cannot be chosen arbitrarily high.

To check the consistency of the method, we have employed polynomials ranging
from second degree to fifth degree, verifying that the results are compatible
within the range of the errors introduced by the incomplete knowledge of the
QCD correlate and other inputs of the calculation. We also have checked
explicitly that a smooth continuum contribution had no influence on the result.

Our approach to suppress the continuum has been tested previously in the
calculation of the heavy quark masses using analogue sum rules for the vector
current correlate where there exists more experimental information on the
continuum. In the calculation of the charm quark mass, using the same
polynomial method, the continuum, known from the BES II collaboration
\cite{bes}, was shown to have practically no influence on the predicted mass
\cite{PENARROCHASCHILCHER}. Employing the same technique, a very accurate
prediction of the bottom quark mass was also obtained using the experimental
information of the upsilon system \cite{PENARROCHABORDESSCHILCHER}.

After these general considerations we proceed with the analytical calculation.
The integrals that we have to evaluate on the right hand side of the sum rule,
equation (\ref{SR}), are
\begin{equation}
J(p,k)=\frac{1}{2\pi i}\oint_{\left\vert s\right\vert =s_{0}}s^{p}\left(
\ln\frac{-s}{\mu^{2}}\right)  ^{k}ds, \label{INTEG}%
\end{equation}
for $k=0,1,2,3$ and $p=-6,-5,..,n+1$. These integrals can be found e.g. in
reference \cite{PPSS}. After integration, equation (\ref{SR}) yields the sum
rule
\begin{align}
M_{D_{q}}^{4}\,\,f_{D_{q}}^{2}P(M_{D_{q}}^{2})  &  =\,m_{c}^{2}(\mu)(m_{c}%
(\mu)+m_{q}(\mu))^{2}\label{SR2}\\
&  \times\sum_{p=0}^{n}\sum_{i=0}^{2}\sum_{j=-1}^{6}\sum_{k=0}^{3}a_{p}\left(
\frac{\alpha_{s}(\mu)}{\pi}\right)  ^{i}\tilde{A}_{j,k}^{(i)}\,\,m_{c}%
^{2j}(\mu)\,\,J(p-j,k)\nonumber\\
&  +\text{ non-perturbative terms}\nonumber
\end{align}
where, for brevity, we have not written down the non-perturbative terms
explicitly. The contribution of the continuum is neglected as explained above although the
continuum threshold is considered in the determination of the coefficients, $a_p$
of the polynomials (\ref{POLY2}).

Plugging the theoretical and experimental inputs (physical threshold, quarks
and meson masses, condensates and strong coupling constant) into the sum rule,
we obtain the decay constant $f_{D_{q}}$ for various values of the degree $n$
of the polynomial and various values of $s_{0}$. Given the correct QCD
asymptotic correlate and the correct hadron continuum, the calculation of
the decay constant should, of course, not depend either on $s_{0}$ or on the
degree $n$ of the polynomial in the sum rule (\ref{SR}). Accordingly, for a
given $n$ we choose the flattest region in the curve $f_{D_{q}}(s_{0})$ to extract
our prediction for the decay constant. To be specific we choose the point of
minimal slope. On the other hand, for different polynomials, the value of
$f_{D_{q}}$, extracted in this way, could differ from each other as the
cancellation of the continuum may be incomplete or the QCD expansion may not
be accurate enough. We find, however, practically the same results for all our
polynomials. This additional stability is truly remarkable as the coefficients
of the polynomials of order $n=2,3,4$ and $5$ are completely different and the
respective predictions are based on completely different superpositions of
finite energy moment sum rules. This extended consistency leads us to attach
great confidence in our numbers and associated errors.

\section{Results}

We calculate the decay constants for the $D$ and $D_{s}$ heavy mesons. In the
first case we take $m_{q}=0$ everywhere. In the second case we retain
$m_{q}=m_{s}\neq0$ in the factor $(M_{c}+m_{q})^{2}$ in front of the
correlation function only. Further terms in the power series in $m_{s}^{2}/s$
in (\ref{PERT2}) are completely negligible for the integration radii $s_{0}$
we use in the calculation.

The experimental and theoretical inputs are as follows. The physical threshold
$s_{\mathrm{phys.}}$ is the squared mass of the lowest lying resonance in the
$c\overline{q}$ channel. For $q$ being the light quark $u$, we have:%

\begin{equation}
s_{\mathrm{phys.}}=M_{D}^{2}=3.493\text{ \ GeV}^{2} \label{BU}%
\end{equation}
whereas the continuum threshold $s_{\mathrm{cont.}}$ is taken from the next
intermediate state $D\pi\pi$ in an s-wave $I=\frac{1}{2}$, i. e.
\[
s_{\mathrm{cont.}}=\left(  M_{D}+2m_{\pi}\right)  ^{2}=4.575\,\,\mathrm{GeV}%
^{2}\,.
\]

For $q$ being the strange quark we take:
\begin{equation}
s_{\mathrm{phys.}}=M_{D_{s}}^{2}=3.873\text{ \ GeV}^{2} \label{BS}%
\end{equation}
The continuum threshold starts in this case at the value:
\[
s_{\mathrm{cont.}}=\left(  M_{D_{s}}+2m_{\pi}\right)  ^{2}%
=5.009\,\,\mathrm{GeV}^{2}.
\]

On the theoretical side of the sum rule we take the following inputs. The
strong coupling constant at the scale of the electroweak $Z$ boson mass
\cite{BETHKE}
\begin{equation}
\alpha_{s}(M_{Z})=0.118\pm0.003 \label{ALPHA}%
\end{equation}
that is run down to the renormalization scale using the four loop formulas of
reference \cite{SANTAMARIA}. For the quark and gluon condensates (see for
example \cite{JAMINLANGE}) and the mass of the strange quark \cite{PDG} we
take:
\begin{align}
&  <\overline{q}q>(2\,\mathrm{GeV})\,=\,(-267\,\pm\,17\,\,\mathrm{MeV}%
)^{3},\nonumber\\
&  <\frac{\alpha_{s}}{\pi}\,G\,G>\,=\,0.024\,\pm\,0.012\,\,\mathrm{GeV}%
^{4},\nonumber\\
&  <\overline{q}\sigma Gq>\,=\,m_{0}^{2}\,<\overline{q}q>,\,\,\,\mathrm{with}%
\,\,\,m_{0}^{2}\,=\,0.8\,\pm0.2\,\mathrm{GeV},\nonumber\\
&  m_{s}(2\,\mathrm{GeV})\,=\,120\,\pm50\,\mathrm{MeV},\nonumber\\
&  <\overline{s}s>\,=\,(0.8\pm0.3)<\overline{q}q>. \label{COND}%
\end{align}

As discussed above, we fix the renormalization scale to be $\mu=m_{c}(m_{c})$.
We use a reasonable variation of $\mu$ to analyze the corresponding
uncertainty in our final result. Finally, for the charm quark, we take the value $m_{c}(m_{c})\,=\,1.25\,\pm
0.10\,\mathrm{GeV}$ obtained by similar techniques
\cite{PDG,PENARROCHABORDESSCHILCHER} which is in a generally accepted range.

In order to calculate the decay constant for the pseudoscalar meson $D$, we
proceed in the way described above. We compute$\ f_{D}$ as a function of
$s_{0}$ with the four different sum rules (\ref{SR}) corresponding to
$n=2,3,4,5$. The results, plotted in Fig. 1 show remarkable stability
properties. We define the optimal value of $s_{0}$ as the center of the
stability region (represented by a cross in Fig.1) where the first and/or
second derivative of $f_{D}(s_{0})$ vanishes. At these values of $s_{0}$ we
obtain the following consistent results:
\begin{align}
f_{D}  &  =176\text{ \ \ MeV for \ \ }n=2\label{POLYRESULT}\\
f_{D}  &  =177\text{ \ \ MeV for \ \ }n=3,4,5\nonumber
\end{align}

\begin{figure}[th]
\centerline{\psfig{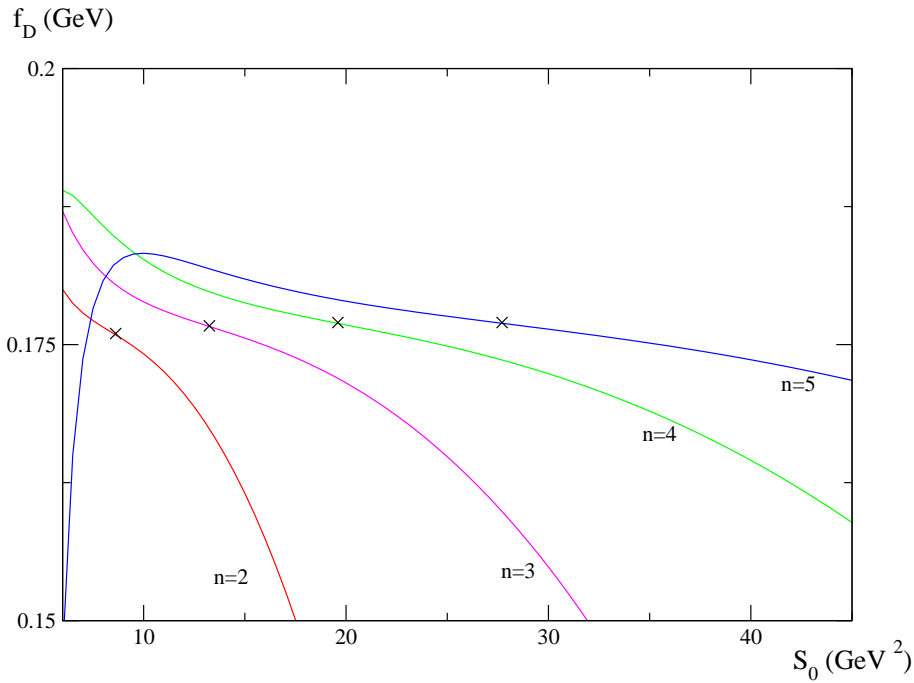}}
\caption{Decay constant $f_{D}$ as a function of the integration radius $s_{0}
$ for $m_{c}(m_{c})=1.25 \, \mathrm{GeV}$. The lines represent the sum rule
(\ref{SR}) for several choices of the polynomial $P_{n}(s)$.}%
\end{figure}

Notice from Fig. 1 that for the fifth degree polynomial (n=5) there is a
stability region of about $20\,\,{\mathrm GeV}^{2}$ around $s_{0}$, where the decay
constant changes by less than three$\,$ percent. From this change we estimate
a conservative error inherent to the method of $\pm3\,\mathrm{MeV}$.

Other sources of errors arising in the calculation of $f_{D}$ are the quark
condensates which affect the result by $\pm5\,\mathrm{MeV,}$ and the charm
mass which, in the range given above, produces a variation in the decay
constant of $-19,+9$ $\mathrm{MeV}$. This is one of the main source of
uncertainty in the final result (see table 1). 
Considering the following results of the perturbative $\alpha_{s}$ expansion:
\begin{align}
\text{one-loop calculation}  &  \text{: \ \ \ }f_{D}^{\left(  0\right)
}=142\text{ \ MeV}  \nonumber\\
\text{two-loop calculation}  &  \text{: \ \ \ }f_{D}^{\left(  1\right)
}=162\text{ \ MeV}\label{ASYMPRESULT}\\
\text{three-loop calculation}  &  \text{: \ \ \ }f_{D}^{\left(  2\right)
}=177\text{ \ MeV}  \nonumber
\end{align}
we also take as a source of uncertainty
the contribution of order $\alpha_{s}^{2}$, which amounts ten percent of the
result. This yields an asymptotic error of $\pm15$ \textrm{MeV }%

We point out that the convergence of the asymptotic series in the present
calculation of the decay constant of the $D$ meson is worse than the one we
found for the $B$ meson \cite{PENARROCHABORDESSCHILCHER2}. Finally we have
considered the dependence on the renormalization scale in the range $\mu
\,\in\,\left[  2,6\right]  \,\mathrm{GeV}$. The error associated to this
change in $\mu$ is roughly related to the convergence of the asymptotic
expansion and therefore it is not considered as an additional one. Other
errors due to the QCD side of the sum rule, higher order terms in $m_{c}%
^{2}/s$ and the error on $\alpha_{s}(m_{Z})$, are negligible in comparison.

\begin{table}[ptb]%
\begin{align*}%
\begin{array}
[c]{|l|l|}\hline
\; \;\; \;\; \;\; \;\; \;\; \; \;\; \;\; \; \; \mathrm{Inputs} & f_{D}=177 \;
\mathrm{MeV}\\\hline
m_{c}(m_{c})\,=1.25\,\pm0.10\,\ \mathrm{GeV} & -19+9 \, \mathrm{MeV}\\
<\overline{q}q>(2\,\mathrm{GeV})\,=\,(-267\,\pm\,17\,\,\mathrm{MeV})^{3} &
\pm\ 5 \mathrm{MeV}\\
<\frac{\alpha_{s}}{\pi}\,G\,G>\,=\,0.024\,\pm\,0.012\,\,\mathrm{GeV}^{4} &
\pm\, 1 \, \mathrm{MeV}\\
<\overline{q}\sigma Gq>\,=\left(  \,0.8\,\pm0.2\,\mathrm{GeV}\right)
^{2}<\overline{q}q>\, & \pm\, 1 \, \mathrm{MeV}\\
\alpha_{s}\left(  m_{c}\right)  =0.335 \pm0.010 & \pm\ 3 \, \mathrm{MeV}%
\\\hline
\end{array}
\end{align*}
\caption{Uncertainties on $f_{D}$ from the uncertainties on the theoretical
parameters}%
\end{table}

From this analysis of errors, we finally quote the following result for the
decay constant of $D$-meson
\begin{equation}
f_{D}\,=\,177\,\pm\,14\,(\mathrm{inp.})\,\pm15\left(\text{\textrm{asymp.}}\right)  
\,\pm\,3\,(\mathrm{meth.}) \,\,\mathrm{MeV}.
\label{RESULTFB}%
\end{equation}

The first error comes from the inputs of the computation, the second to the truncated QCD
theory whereas the last one is due to the method itself.

Proceeding in the same fashion, but keeping the mass of the strange quark in
the overall factor and the order $m_{s}/m_{c}$ in the one loop contribution,
we find the decay constant for the $D_{s}$ meson,
\begin{align}
\text{one-loop calculation}  &  \text{: \ \ \ }f_{D_{s}}^{\left(  0\right)
}=163\text{ \ MeV}\nonumber\\
\text{two-loop calculation}  &  \text{: \ \ \ }f_{D_{s}}^{\left(  1\right)
}=188\text{ \ MeV}\label{ASYMPRESULTDS}\\
\text{three-loop calculation}  &  \text{: \ \ \ }f_{D_{s}}^{\left(  2\right)
}=205\text{ \ MeV}\nonumber
\end{align}
and including the analysis of uncertainties we find:
\[
f_{D_{s}}\,=\,205\,\pm\,14\,(\mathrm{inp.})\,\pm 17 \left(\text{\textrm{asymp.}}\right)  
\,\pm\,3\,(\mathrm{meth.})%
)\,\,\mathrm{MeV}.
\]
(see Fig. 2)

\begin{figure}[th]
\centerline{\psfig{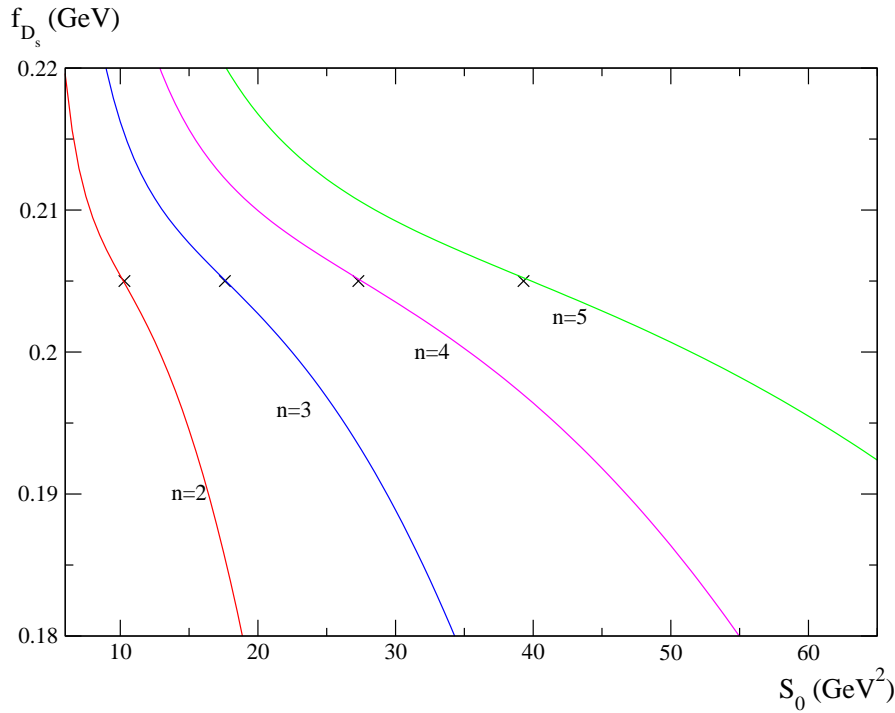}}
\caption{Decay constant $f_{D_{s}}$ as a function of the integration radius
$s_{0}$ for $m_{c}(m_{c})=1.25\,\mathrm{GeV}$. The lines represent the sum
rule (\ref{SR}) for several choices of the polynomial $P_{n}(s)$.}%
\end{figure}

In the analysis of theoretical errors the only new ingredient is the
uncertainty coming from the strange quark mass which turns out to be negligible.

The ratio of the decay constants $f_{D_{s}}$ and $\ f_{D}$ (which would be $1
$ in the chiral limit) is of special interest. We find:%

\begin{equation}
\frac{f_{D_{s}}}{f_{D}}\,=\,1.16\,\pm\,0.01\,(\mathrm{inp.})\,\pm
\,0.03\,(\mathrm{meth.}).
\end{equation}
in complete agreement with lattice calculations. The uncertainties due to the
theoretical inputs are correlated, so that the final error is very small.

\section{Conclusions}

In this note we have computed the decay constant of $D_{q}$-mesons for $q$
being either the strange or the $u$ or $d$ massless quarks. We have employed a
judicious combination of moments in QCD finite energy sum rules in order to
minimize the shortcomings of the available experimental data. On the
theoretical side of the pseudoscalar two-point function, we have used in the
perturbative piece an expansion up to three loops in the strong coupling
constant and up to order $\left(  m_{c}^{2}/s\right)  ^{7}$ in the mass
expansion and in the non-perturbative piece we considered condensates up to
dimension six including the $\alpha_{s}$ correction in the leading term.
Instead of the commonly adopted pole mass of the bottom quark, we use the
running mass to improve convergence of the perturbative series.

In the sum rule, the contour integration of the asymptotic part is performed
analytically. This particular fact differs from other computations based on
sum rules where the asymptotic QCD is integrated along a cut of the two-point
function starting at the pole mass squared. The latter way to proceed is
problematic when loop corrections are included and the complete analytical QCD
expression along the cut is not known. In this approach QCD has to be
extrapolated from low energy to high energy \cite{JAMINLANGE}. We also differ
from many other sum rule calculation in that we do not require two unrelated
sum rules to determine a duality point via an intercept of the curves
$f_{D}(s_{0})$.

Our results are very sensitive to the value of the running mass, giving most
of the theoretical uncertainty. They also turn out to be sensitive to
variations of the renormalization scale $\mu.$ The uncertainties of other
theoretical parameters like quark condensates and coupling constant are
less important. Adding quadratically the different estimated errors we have
the final results%
\begin{eqnarray}
f_{D}\,  &  = & \,177\,\pm\,21\,\ \mathrm{MeV} \nonumber \\
f_{D_{s}}\,  &  = & \,205\,\pm\,22\,\,\mathrm{MeV}%
\label{FINAL}
\end{eqnarray}

Comparing (\ref{FINAL}) with other results in the literature, our results agree within the
error bars with the ones obtained using sum rule methods
\cite{PENINSTEINHAUSER,NEUBERT}. However, compared with lattice computations
\cite{APE,UKQCD} they are a bit lower.

\bigskip

\section*{Appendix}

For convenience of the reader we list in this appendix the first few
polynomials emerging from relations (\ref{POLYNORM},\ref{POLY2}). From the
second condition, namely (\ref{POLY2}), it is easy to realize that the set of
polynomials $P_{n}(s)$ are n-degree orthogonal polynomials in the interval of
the variable $s\,\in\,[s_{\mathrm{cont.}},s_{0}]$. With the
normalization condition (\ref{POLYNORM}) (adopted to emphasize the
contribution of the lowest lying resonance in the sum rule) they are related
to the Legendre polynomials $\mathcal{P}_{n}(x)$ in the interval of the
variable $x\,\in\,[-1,1]$ as follows:
\begin{equation}
P_{n}(s)\,=\,\frac{\mathcal{P}_{n}\left(  x(s)\right)  }{\mathcal{P}%
_{n}\left(  x(M_{D_{q}}^{2})\right)  } \label{LEGENDRE}%
\end{equation}
Where the variable $x(s)$ is:
\[
x(s)\,=\,\frac{2s\,-\,(s_{0}+s_{\mathrm{cont.}})}%
{s_{0}-s_{\mathrm{cont.}}}%
\]
Obviously $x(s)\,\in\,[-1,1]$ when $s\,\in\,[s_{\mathrm{cont.}},s_{0}]$.

The explicit form of these polynomials is well known and can be found, for
instance, in \cite{SANSONE}. Nevertheless, for sake of completeness, we quote
here the ones we have used in the calculation.
\begin{align*}
&  \mathcal{P}_{2}\left(  x(s)\right)  \,=\,\frac{1}{2}(3x^{2}-1),\\
&  \mathcal{P}_{3}\left(  x(s)\right)  \,=\,\frac{1}{2}(5x^{3}-3x),\\
&  \mathcal{P}_{4}\left(  x(s)\right)  \,=\,\frac{1}{8}(35x^{4}-30x^{2}+3),\\
&  \mathcal{P}_{5}\left(  x(s)\right)  \,=\,\frac{1}{8}(63x^{5}-70x^{3}+15x).
\end{align*}

Finally, in Fig. 3 and in order to appreciate the suppression of the
experimental physical continuum data in the sum rule, we have plotted the form
of the polynomials $P_{n}(s)$ for n=2,3,4,5 at the stability values of $s_{0}$
used in the calculation of $f_{D}$.

\begin{figure}[tb]
\centerline{\psfig{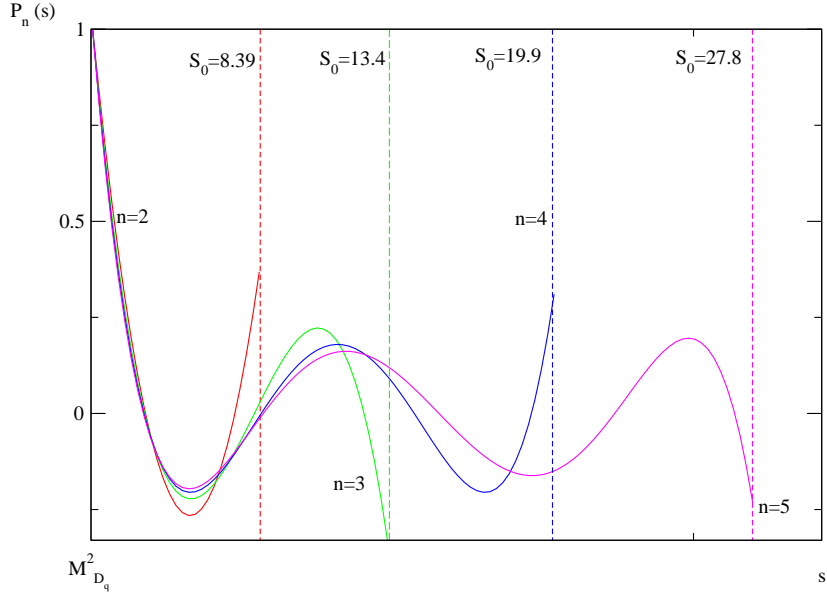}}
\caption{Polynomials $P_{n}(s)$ obtained from conditions (\ref{POLYNORM}%
,\ref{POLY2}) taken at the stability values of $s_{0}$ in the calculation of
$f_{B}$.}%
\end{figure}

\end{document}